\documentclass[preprint,cite]{epl}
\usepackage{amsmath,amssymb}

\newcommand{\tw}{t_\mathrm{w}}

\title{Length scale dependence of dynamical heterogeneity in a
colloidal fractal gel} \shorttitle{Dynamical heterogeneity in a gel}
\author{Agn\`{e}s Duri and Luca Cipelletti\inst{1}}
\shortauthor{Duri \etal} \institute{
 \inst{1} Laboratoire des Collo\"{\i}des, verres et Nanomat\'{e}riaux
(UMR CNRS-UM2 5587), cc26, Universit\'{e} Montpellier 2, 34095
Montpellier Cedex 5, France}
 {Glass transitions}
 {Gels and Sols}
\begin{document}
\maketitle
\begin{abstract}
We use time-resolved dynamic light scattering to investigate the
slow dynamics of a colloidal gel. The final decay of the average
intensity autocorrelation function is well described by
$g_2(q,\tau)-1 \sim \exp[-(\tau/\tau_\mathrm{f})^p]$, with
$\tau_\mathrm{f} \sim q^{-1}$ and $p$ decreasing from 1.5 to 1
with increasing $q$. We show that the dynamics is not due to a
continuous ballistic process, as proposed in previous works, but
rather to rare, intermittent rearrangements.  We quantify the
dynamical fluctuations resulting from intermittency by means of
the variance $\chi(\tau,q)$ of the instantaneous autocorrelation
function, the analogous of the dynamical susceptibility $\chi_4$
studied in glass formers. The amplitude of $\chi$ is found to grow
linearly with $q$. We propose a simple --yet general-- model of
intermittent dynamics that accounts for the $q$ dependence of both
the average correlation functions and $\chi$.
\end{abstract}


Soft matter systems where the constituents are packed at high
volume fraction or interact strongly have a dynamical behavior
reminiscent of that of molecular glasses~\cite{Donth2001}: they
exhibit very slow relaxations, non-exponential response or
correlation functions, history-dependent dynamics, and dynamical
heterogeneity~\cite{CipellettiJPCM2005}. However, soft glassy
systems may also exhibit peculiar dynamical features not found in
hard glasses. An example is provided by low volume fraction
colloidal gels resulting from the aggregation of strongly
attractive particles. In these gels, the decay of the intensity
correlation function $g_2(q,\tau)-1$ measured by dynamic light
scattering is steeper than exponential and the relaxation time,
$\tau_\mathrm{f}$, has an anomalous $q^{-1}$ dependence on the
scattering vector~\cite{LucaPRL2000}. This ``compressed''
exponential, ballistic-like dynamics has to be contrasted with the
stretched exponential relaxations and the diffusive behavior
($\tau_\mathrm{f} \sim q^{-2}$) usually found in molecular
systems~\cite{Donth2001}. Quite intriguingly, this unusual
dynamics is not restricted to dilute colloidal gels, but has been
observed recently in a large variety of soft systems with both
attractive and repulsive interactions
~\cite{RamosPRL2001,LucaFaraday2003,BellourPRE2003,BandyopadhyayPRL2004,HardenPRL2006,MochriePRL2006,RobertEPL2006}.

For colloidal gels, it has been proposed that the dynamics be due
to the continuous evolution of strain fields set by dipolar
sources of internal
stress~\cite{LucaPRL2000,BouchaudEPJE2001,LucaFaraday2003}. At a
microscopic level, stress is presumably accumulated by changes in
the local structure driven by the formation of new bonds and/or
bond breaking, which constitute elementary steps in the direction
of a more stable, compact structure. More generally, a similar
continuous evolution of dipolar strain fields has been invoked to
explain the ballistic-like dynamics of other systems exhibiting a
compressed exponential relaxation of
$g_2$~\cite{RamosPRL2001,LucaFaraday2003,BellourPRE2003,BandyopadhyayPRL2004,HardenPRL2006,MochriePRL2006,RobertEPL2006}.

Time-resolved light scattering experiments performed on similar
gels~\cite{LucaJPCM2003} have suggested that the dynamics be
temporally heterogeneous, thus challenging the arguments of
refs.~\cite{LucaPRL2000,BouchaudEPJE2001,LucaFaraday2003} that are
based on a continuous dynamics~\cite{noteBouchaudIntermittency}.
However, these experiments were performed on samples at much
higher concentration ($\varphi \sim 0.1$ as opposed to
$10^{-3}-10^{-4}$) and in the multiple scattering regime, where
the dynamics in probed on much shorter length scales (less than a
particle size as opposed to hundreds of particle sizes). Whether
their outcome might be extrapolated to single scattering
experiments on diluted gels at low $q$ is thus still unclear,
especially since time-resolved light scattering measurements in
the same low-$q$ range as that of ref.~\cite{LucaPRL2000} failed
to evidence any significant temporal
heterogeneity~\cite{ManleyPersonalComm}. Clarifying the nature and
the physical origin of the dynamics of the gels is particularly
important, given the widespread occurrence of similar dynamics in
soft glassy materials.

In this Letter, we tackle these issues by studying the dynamics of
strongly attractive colloidal gels in a range of $q$ vectors about
one decade higher than in~\cite{LucaPRL2000} and using
time-resolved dynamic light scattering. We find that at all $q$
the final relaxation of the average intensity correlation function
$g_2(q,\tau)-1$ is well fitted by a compressed exponential with
the same $q^{-1}$ dependence of $\tau_\mathrm{f}$ as observed
previously~\cite{LucaPRL2000}, and with a compressing exponent
$p>1$. While at the lowest $q$ $p\approx 1.5$, the exponent
unexpectedly decreases with increasing $q$, eventually approaching
one. We demonstrate that the dynamics is intermittent and quantify
the resulting fluctuations of $g_2$ by means of a ``multipoint''
correlation function $\chi$ analogous to the dynamical
susceptibility $\chi_4$ introduced in simulations of glass
formers~\cite{FranzJPCM2000,LacevicJChemPhys2003,BerthierPRE2004}.
We find that $\chi$ increases linearly with $q$ and introduce a
general model of intermittent dynamics that accounts for both the
average dynamics and its temporal fluctuations.

The gels are made of polystyrene particles of radius $10$ nm
suspended in a buoyancy matching mixture of H$_{2}$O and D$_{2}$O
(45/55 by volume). Particles are mixed with a MgCl$_2$ solution in
order to induce aggregation in the DLCA
regime~\cite{CarpinetiPRL1992}. The final particle volume fraction
and salt concentration are $\varphi = 6 \times 10^{-4}$ and 10 mM,
respectively. A gelled structure is obtained after about 2 hours,
formed by interconnected fractal clusters of radius $R_\mathrm{c}
\approx 10~\mu\mathrm{m}$, as revealed by static light scattering
~\cite{CarpinetiPRL1992,LucaPRL2000}. The dynamics of the gel slows
down with age~\cite{LucaPRL2000}. To prevent significant aging
during the experiment, we focus on a time window of duration $T_{\rm
exp} = 20000$ sec starting at $\tw = 280\,000~\textrm{sec}$, where
$\tw =0$ is the time when the gel is formed.

The gel dynamics is measured by using a charge-coupled device
(CCD) camera-based light scattering apparatus similar to that
described in~\cite{LucaRSI1999}, slightly modified to access
larger $q$ vectors. The dynamics is measured simultaneously at
several $q$'s ($0.4~\mu\mathrm{m}^{-1} \le q \le
5.5~\mu\mathrm{m}^{-1}$), corresponding to length scales
intermediate between the particle and the cluster size. In order
to access both the average dynamics and its temporal fluctuations,
we use the time-resolved correlation
scheme~\cite{LucaJPCM2003,DuriPRE2005}. The degree of correlation
between pairs of images of the speckle pattern scattered at time
$\tw$ and $\tw + \tau$ is calculated according to $c_I(\tw,\tau,q)
= G_2(\tw,\tau)/
\left( \left <I_p(\tw) \right
>_p \left <I_p(\tw + \tau ) \right >_p \right) - 1$, where
$G_2(\tw,\tau) = \left< I_p(\tw)I_p(\tw+\tau) \right >_p$ and
$I_p(t)$ is the scattered intensity at pixel $p$ and time $t$.
$\left < \cdot \cdot \cdot \right>_p$ is an average over pixels
corresponding to the same magnitude of $\mathbf{q}$ but different
azimuthal orientations. The intensity autocorrelation function is
$g_2(q,\tau)-1 = \overline{c_I(\tw,\tau,q)}$,  where
$\overline{\cdot\cdot\cdot}$ indicates a time average over $T_{\rm
exp}$. Dynamical fluctuations are quantified by the temporal
variance of $c_I$ at fixed $\tau$ and
$q$~\cite{LucaJPCM2003,DuriPRE2005}, as discussed in more detail in
the following.

\begin{figure}
\onefigure{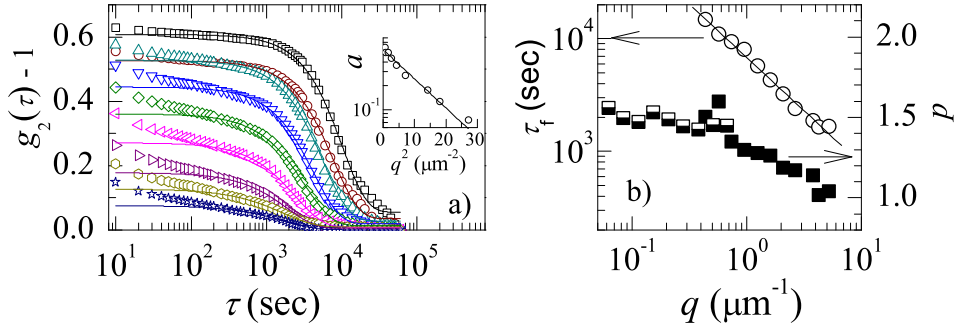} \caption{a): $g_2(q,\tau)-1$ for a gel at
$\varphi = 6 \times 10^{-4}$. From top to bottom, $q$ varies from
0.74 to 5.22 $\mu\mathrm{m}^{-1}$. The lines are compressed
exponential fits to the final relaxation of $g_2-1$. Inset:
semilogarithmic plot of the plateau height $a$ \textit{vs.} $q^2$.
The line is an exponential fit. b): $q$ dependence of the relaxation
time $\tau_{\rm f}$ (open circles, left axis) and of the compressing
exponent $p$ (right axis, solid squares: this work; semiopen
squares: ref.~\cite{LucaPRL2000}). The line is a power law fit to
$\tau_{\rm f}$ yielding an exponent $-0.94 \pm 0.03$.} \label{fig1}
\end{figure}

The average dynamics is shown in Fig.~\ref{fig1}a) for several
$q$'s. At all scattering vectors, $g_2(q,\tau)-1$ exhibits an
initial decay, followed by a slightly tilted plateau and a final
relaxation. The initial decay is barely observable due to the
limited frame rate of the CCD camera; it is due to overdamped,
thermally activated fluctuations of the gel
strands~\cite{KrallPRL1998}. The height $a$ of the plateau is
related to the average amplitude, $\delta_{\rm p}$, of these
fluctuations by $a \sim \exp(-q^2\delta^2_{\rm
p}/3)$~\cite{KrallPRL1998}. By fitting $a(q)$ to this Gaussian
form, we find $\delta_{\rm p} = 500 \pm 150$ nm (inset of
Fig.~\ref{fig1}a). The final relaxation is well fitted by a
compressed exponential decay: $g_2(q,\tau)-1 = a
\exp[-(\tau/\tau_{\rm f})^p]$, where $a$, $\tau_{\rm f}$, and $p$
depend on $q$. Figure~\ref{fig1}b) shows the $q$ dependence of $p$
(solid and semiopen symbols) and $\tau_{\rm f}$ (open circles). We
find $\tau_{\rm f}\sim q^{-0.94 \pm 0.03}$, consistently with
previous measurements at lower $q$~\cite{LucaPRL2000}. This
behavior rules out diffusive motion and indicates that, on
average, the particle displacement increases linearly with time.
The compressing exponent $p$ is approximately $1.5$ at the
smallest $q$ probed in this work, in agreement with
ref.~\cite{LucaPRL2000} (semiopen squares) and in analogy with
refs~\cite{RamosPRL2001,LucaFaraday2003,BandyopadhyayPRL2004,HardenPRL2006,RobertEPL2006}.
Surprisingly, however, at larger scattering vectors $p$ decreases
with $q$, finally approaching $p=1$. A similar trend has been
reported very recently for other
systems with analogous dynamics~\cite{
MochriePRL2006,RobertEPL2006}, although its origin remained
unclear. We stress that this behavior is incompatible with the
arguments proposed in refs.~\cite{LucaPRL2000,LucaFaraday2003},
which predict $p=1.5$ regardless of $q$. The model by Bouchaud and
Pitard predicts $p=1.5$ for $q\rightarrow 0$ and $p=1.25$ for
$q\rightarrow \infty$ \cite{BouchaudEPJE2001}. However, in the
intermediate $q$ regime $g_2-1$ should deviate from a compressed
exponential, contrary to our observations.

\begin{figure}
\onefigure{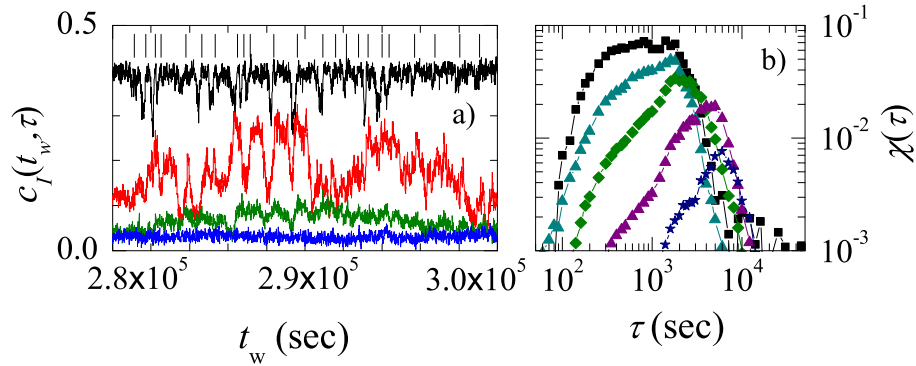} \caption{a): temporal fluctuations of the
instantaneous degree of correlation $c_I$ for
$q=2.07~\mu\mathrm{m}^{-1}$. From top to bottom, $\tau=250, 3000,
8000,~\mathrm{and}~20000~\mathrm{sec}$. The vertical bars above the
top curve indicate the time of individual rearrangement events. b):
dynamical susceptibility $\chi(\tau,q)$ (normalized variance of
$c_I$). From bottom to top, $q =
0.74,~1.24,~2.07,~3.78,~\mathrm{and}~5.22~\mu\mathrm{m}^{-1}$ (same
symbols as in Fig.~\ref{fig1}a). For the sake of clarity, not all
the available curves have been plotted.} \label{fig2}
\end{figure}

To better understand the origin of the gel dynamics, we investigate
time-resolved quantities. Figure~\ref{fig2}a) shows the time
dependence of the instantaneous degree of correlation $c_I$ for $q =
2.07~\mu\mathrm{m}^{-1}$ and for several time lags $\tau$ (similar
results are obtained for all $q$'s). The top curve shows data for
$\tau = 250$ sec, a time lag much smaller than the relaxation time
$\tau_{\rm f} = 3220$ sec; the data are correct for the contribution
of measurement noise as described in Sec. IV c of
ref.~\cite{DuriPRE2005}. Sharp drops of $c_I$ departing from a high
plateau are visible, indicating sudden changes in the speckle
pattern caused by discrete rearrangement events in the gel. The
typical width of the downward spikes correspond essentially to
$\tau$, indicating that the rearrangements are shorter than the
temporal resolution of the experiment (10 sec)~\cite{DuriFNL2005}.
We locate the time of the events by analyzing the drops of $c_I$ for
$\tau=250$ sec and by verifying that similar drops are observed at
the same time for all $q$'s. The vertical bars in Fig.~\ref{fig2}a)
indicate the occurrence of the rearrangements thus identified. We
count 23 events during the time $T_{\rm exp} = 20000$ sec, implying
that the average time between events is $\Delta t  = T_{\rm exp}/23
= 870$ sec. For $\tau = 3000~{\rm sec}\approx \tau_{\rm f}$, much
larger fluctuations of $c_I $ are observed, because the number of
events occurring between two speckles images separated by 3000 sec
may significantly vary, due to the random nature of the events. This
has to be contrasted with the case $\tau = 250$ sec $<< \Delta t$,
where at most one event occurs between two images, yielding the
discrete drops discussed above. For $\tau >> \Delta t$ (two bottom
curves in Fig.~\ref{fig2}a), the fluctuations are again reduced,
because on this time scale many events occur and hence their
(relative) number fluctuations are smaller, by virtue of the Central
Limit theorem.

\begin{figure}
\onefigure{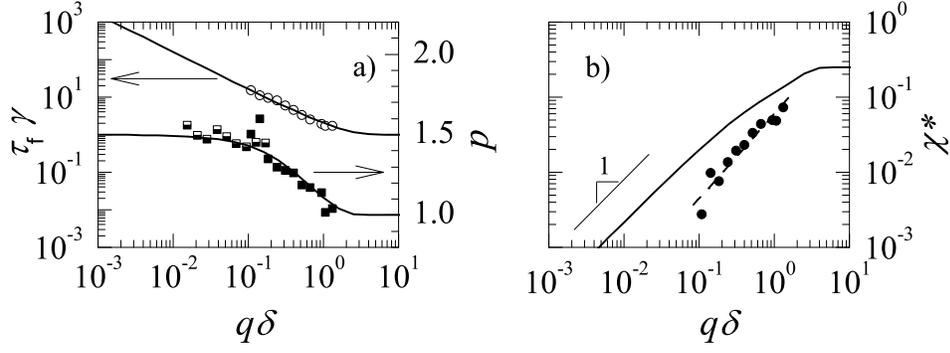} \caption{a): $q$ dependence of the relaxation
time (top line, left axis) and of the compressing exponent (bottom
line, right axis) in reduced variables, as obtained from the model
described in the text. The symbols are the experimental data shown
in Fig.~\ref{fig1}b): a very good agreement with the model is
obtained for $\delta = 250$ nm and $\gamma^{-1} = 960$ sec. b) peak
of the dynamical susceptibility $\chi^*$ $vs$ $q\delta$ for the
model (line) and the experiments (symbols). The dashed line is a
power law fit with an exponent $1.13 \pm 0.11$} \label{fig3}
\end{figure}

We quantify the fluctuations of the dynamics resulting from the
intermittent rearrangements by calculating $\chi(\tau,q)$, the
variance of $c_I$ corrected for the measurement noise
contribution~\cite{DuriPRE2005}. As discussed
in~\cite{
DuriPRE2005}, this quantity is the analogous in light scattering
experiments of the dynamical susceptibility $\chi_4$ much studied in
theoretical and numerical works on glass formers and
gels~\cite{FranzJPCM2000,LacevicJChemPhys2003,BerthierPRE2004,deCandiaPhysicaA2005}.
In order to compare data taken at different $q$ vectors, we focus on
the relative fluctuations by normalizing the variance of $c_I$ with
respect to the amplitude of the final decay of $g_2-1$:
$\chi(\tau,q) = a(q)^{-2}\left [
\overline{c_I(\tw,\tau,q)^2}-\overline{c_I(\tw,\tau,q)}^{\,2}\right
]$. Results for some representative $q$ vectors are shown in
Fig.~\ref{fig2}b): for all $q$, $\chi$ has a peaked shape, the peak
position corresponding approximately to the decay time of the
average correlation function, where the fluctuations of $c_I$ are
largest as observed in Fig.~\ref{fig2}a). This peaked shape is
analogous to that reported in previous simulations and experiments
on other glassy
systems~\cite{LacevicJChemPhys2003,DuriPRE2005,MayerPRL2004,deCandiaPhysicaA2005,DauchotPRL2005_2}.
In these systems, fluctuations usually arise from \textit{spatial}
correlations of the dynamics that reduce the number of statistically
independent regions in the system. By contrast, for the gels $\chi$
is dictated by the \textit{temporal} intermittency of the
rearrangement events, since they affect simultaneously the whole
scattering volume, as found by directly measuring the spatial
correlation of the slow dynamics~\cite{DuriSpatCorr}. The length
scale dependence of the fluctuations, quantified by the height of
the peak, $\chi^*(q)$, has a very surprising behavior. As shown in
Fig.~\ref{fig3}b), we find that $\chi^*$ increases strongly with $q$
($\chi^* \sim q^{1.13 \pm 0.11}$). To our knowledge, no other
experimental data on 3D systems are available; however we point out
that the gel behavior is in contrast with measurements on a 2D
granular system, where the peak of $\chi_4$ showed virtually no $q$
dependence~\cite{DauchotPRL2005_2}.

The intermittent dynamics shown here is in stark contrast with the
continuous ballistic motion assumed in
\cite{LucaPRL2000,LucaFaraday2003,BouchaudEPJE2001}. In the
following, we develop a simple yet general model of dynamic light
scattering from an intermittent dynamical process, aiming at
rationalizing both the average correlation function and its
fluctuations. We assume the slow dynamics to be due to a series of
discrete rearrangement events, which we take --for simplicity-- to
have equal amplitude and to be instantaneous. The degree of
correlation is then a function $h$, to be determined, of $q$ and the
number $m$ of events occurring between $\tw$ and $\tw+\tau$:
$c_I(\tw,\tau,q) = h[m(\tw,\tau),q]$. Within this scenario, the
fluctuations of $c_I$ are due to fluctuations of the number of
events actually occurring during any given interval
$[\tw,\tw+\tau]$. We calculate the average dynamics and its
fluctuations according to
\begin{eqnarray}
g_2(\tau,q)-1 = \sum_{n=0}^{\infty} P_{\tau}(n)h(n,q) \label{eq1}
\end{eqnarray}
\begin{eqnarray} \chi(\tau,q) = \sum_{n=0}^{\infty}
P_{\tau}(n)[h(n,q)-(g_2(\tau,q)-1)]^2 \,, \label{eq2}
\end{eqnarray}
where $P_{\tau}(n)$ is the probability that $n$ events affect the
scattering volume during a time span $\tau$. In writing
Eq.~(\ref{eq2}), we have taken into account that the dynamics is
spatially correlated over a length scale much larger than the size
of the scattering volume, so that fluctuations are not reduced by
averaging the collected signal over several dynamically independent
regions.

In order to calculate $g_2$ and $\chi$ from Eqs.~(1,2), we need
expressions for $P_{\tau}$ and $h$. For the former, we choose for
simplicity a Poisson law: $P_{\tau}(n) = \exp(-\gamma \tau)(\gamma
\tau)^n/n!$, corresponding to rearrangement events that are random
in time and affect the scattering volume at an average rate
$\gamma$. For the latter, we write $h(n,q)$ in terms of the
displacement field generated by $n$ rearrangement events. By
introducing $\Delta \mathbf{R}$, the particle displacement due to
one single event, one has $h(n,q) = <\exp(-in^{\alpha}\mathbf{q}
\cdot \Delta \mathbf{R})>$~\cite{Berne1976}, where the average is
taken over all possible orientations of $\mathbf{q}$ and over all
particles. The exponent $\alpha$ is expected to be $\le 1$:
$\alpha = 0.5$ would correspond to diffusive-like dynamics where
the displacement grows as the square root of the number of events,
while $\alpha=1$ would lead, on average, to a ballistic-like
motion. For the gels, we expect $\alpha=1$, because for the
average dynamics $q \propto \tau_{\rm f}^{-1}$. The correlation
left after $n$ events is then $h(n,q) =
<\exp(-in\mathbf{q}\cdot\Delta \mathbf{R})> = \int PDF(\Delta
\mathbf{R}) \exp(in\mathbf{q}\cdot\Delta \mathbf{R})
\mathrm{d}\Delta \mathbf{R}$ , where $PDF(\Delta \mathbf{R}) $ is
the probability distribution function (PDF) of the particle
displacements~\cite{Berne1976}. By assuming that the displacement
field induced by one single event is that due to the long range
elastic deformation of the gel under the action of dipolar
stresses~\cite{BouchaudEPJE2001,LucaFaraday2003} and using
arguments similar to those developed in
ref.~\cite{LucaFaraday2003}, one has $h(n,q) =
\exp[-(qn\delta)^{\beta}]$. Here $\delta$ is the typical
displacement of the particles due to one single event and
$\beta=1.5$ corresponds to a PDF with a $\Delta R^{-2.5}$ power
law right tail, as predicted for dipolar stress sources randomly
scattered in space~\cite{BouchaudEPJE2001,LucaFaraday2003}.

We insert the above expressions for $P_{\tau}$ and $h$ in
Eqs.~(\ref{eq1},\ref{eq2}), using $\alpha=1, \beta=1.5$, and
calculate both the average dynamics and its fluctuations. The
resulting $g_2(\tau,q)-1$ is very well approximated by a compressed
exponential decay, in agreement with the experimental data. We show
in Fig.~\ref{fig3}a) the compressing exponent $p$ and the
dimensionless relaxation time $\gamma \tau_{\rm f}$ issued from the
fit of the model, as a function of the dimensionless scattering
vector $q\delta$ (lines). For $q\delta \rightarrow 0$, we find $p
\rightarrow \beta = 1.5$. At larger scattering vectors, $p$
decreases approaching one. In all the $q$ regime where $p>1$,
$\gamma \tau_{\rm f} \propto q^{-1}$, a consequence of the choice
$\alpha = 1$ that yields, on average, ballistic-like dynamics. Note
that the fact that the displacement due to successive events adds up
coherently ($\alpha=1$) suggests that they originate from the same
source, i.e. that stress sources are very diluted and their life
time exceeds the relaxation time of $g_2$. Thus, several discrete
events are necessary to fully relax a stress source. For $q\delta>>
1$, where $p$ saturates to one, we find $\tau_{\rm f} = \gamma^{-1}$
independently of $q$. Indeed, when the typical displacement $\delta$
is much larger than the length scale $1/q$ probed by light
scattering, one single event is sufficient to lead to a complete
decorrelation of the scattered light. The only relevant time scale
is then $\gamma^{-1}$, the average time between events. In this
regime $h(n,q) \approx 0$ for $n>0$: the only non vanishing term in
the average correlation function, r.h.s. of Eq.~(1), corresponds to
$n=0$ and $g_2(q,\tau)-1 = \exp(-\gamma \tau)$.

Figure ~\ref{fig3}a) shows also the experimentally determined $p$
and $\tau_{\rm f}$ plotted using dimensionless variables, with
$\delta= 250$ nm and $\gamma = 1.04 \times 10^{-3}$ Hz. With this
choice of the parameters, a very good agreement between the model
and the experiments is found. Remarkably, $\delta$ is of the same
order of magnitude of the amplitude of the fluctuations of the gel
strands due to thermal motion ($\delta_\mathrm{p} = 500$ nm). This
strongly supports the intuitive picture that the rearrangement
events correspond to the formation of new bonds and/or the
breaking of the bonds along the gel network. Since both processes
are ultimately triggered by thermal energy, they would typically
cause particle displacements comparable to those due to the
thermal fluctuations of the gel. The average ballistic dynamics is
thus the result of a slow compaction of the gel that proceeds by
discrete rearrangements, rather than of a continuous process. The
event rate $\gamma$ determined by fitting the model to the
experimental average dynamics corresponds to an average time
between events of 960 sec, in excellent agreement with $\Delta t =
870$ sec obtained directly from the drops of $c_I$ shown in
Fig.~\ref{fig2}a). A further test of the model would be the
observation of the high $q$ limit where $g_2-1 \sim \exp(-\gamma
\tau)$. However, for our gels this regime is not experimentally
accessible, since as $q$ grows the amplitude $a$ of the slow mode
of $g_2-1$ decreases as $\exp(-q^2\delta^2_{\rm p}/3)$ (see
Fig.~\ref{fig1}a), ultimately vanishing for $q\delta_{\rm p} >
q\delta >> 1$.

Having fixed the parameters of the model by comparison with the
experimental average dynamics, we compare the theoretical
predictions for the dynamical susceptibility to the experiments.
We find that $\chi$ has the same peaked shape as for the
experiments; the $q$ dependence of the height of the peak is shown
in Fig.~\ref{fig3}b) (line) together with the experimental points
(filled circles). The model captures correctly the nearly linear
growth of $\chi*$ with $q$ 
and the order of magnitude of the dynamical fluctuations, although
it overestimates them by about a factor of two. Given the
simplicity of the assumptions and the fact that no adjustable
parameter has been introduced specifically for $\chi$, this
agreement is quite remarkable, showing that the model captures the
essential physics of the slow dynamics of the gel. Interestingly,
the $\chi^* \sim q$ growth at low $q$ and the saturation regime at
large $q$ are similar to asymptotic predictions for
$\chi_4(\tau,q)$ in glass formers proposed very recently by
Chandler et al. Indeed, in ref.~\cite{Chandlercondmat0605084} a
saturation regime at large $q$ and a $\chi_4 \sim q^2$ scaling for
$q\rightarrow0$ are identified, the $q^2$ rather that $q$
dependence being the consequence of diffusive rather than
ballistic-like dynamics.

In conclusion, we have shown that the slow dynamics of colloidal
gels is a highly discontinuous process, due to intermittent,
sudden rearrangement events. We have proposed a simple model of
intermittent dynamics that captures correctly the $q$ dependence
of both the average dynamics and dynamical fluctuations. Although
some aspects of our model and their physical interpretation are
certainly specific to the dilute colloidal gels studied here, its
general features may be relevant also to other soft systems. In
particular, more time-resolved experiments will be needed to test
whether the underlying picture of intermittent rearrangements may
apply also to the numerous soft systems whose average dynamics is
similar to that of the gels. This would challenge the currently
accepted interpretation in terms of a continuous ballistic
process~\cite{RamosPRL2001,LucaFaraday2003,BellourPRE2003,BandyopadhyayPRL2004,HardenPRL2006,MochriePRL2006,RobertEPL2006}.

We thank V. Trappe, P. Chaikin, D. Pine, J. P. Bouchaud and L.
Berthier for useful discussions. This work was supported in part by
the European MCRTN ``Arrested matter'' (MRTN-CT-2003-504712) and the
NoE ``SoftComp'' (NMP3-CT-2004-502235), and by CNES
(03/CNES/4800000123) and the Minist\`{e}re de la Recherche (ACI
JC2076). L.C. is a junior member of the Institut Universitaire de
France.


\end{document}